\documentclass[aps,prb,longbibliography,showpacs,twocolumn,superscriptaddress]{revtex4-2}
\usepackage{amsmath,amssymb,amsfonts,bm}
\usepackage{booktabs}
\usepackage{braket}
\usepackage{graphicx}
\usepackage{epstopdf}
\usepackage{dcolumn}
\usepackage{mathrsfs}
\usepackage{bbold}
\usepackage{dsfont}
\usepackage{float}
\usepackage[colorlinks=true,linkcolor=blue,citecolor=blue, urlcolor=blue,bookmarks=false]{hyperref}

\usepackage{tgtermes}
\usepackage{multirow}

\begin{document}
	
\title{Altermagnetism in moir\'e quasicrystals} 

\author{Zi-Ming Wang}
\affiliation{Department of Physics and Chongqing Key Laboratory for Strongly Coupled Physics, Chongqing University, Chongqing 400044,  China}

\author{Rui Chen}
\affiliation{Department of Physics, Hubei University, Wuhan 430062, China}

\author{Dong-Hui Xu}
\email{donghuixu@cqu.edu.cn}
\affiliation{Department of Physics and Chongqing Key Laboratory for Strongly Coupled Physics, Chongqing University, Chongqing 400044, China}
\affiliation{Center of Quantum Materials and Devices, Chongqing University, Chongqing 400044, China}

\date{\today}
	
\begin{abstract}
Altermagnetism is a collinear compensated magnetic phase with zero net magnetization and momentum-dependent spin polarization. Quasicrystalline altermagnetism extends altermagnetism beyond periodic crystals, with noncrystallographic symmetries allowing high-order spin polarization patterns. Here, we propose a symmetry-based approach to realizing quasicrystalline altermagnetism by twisting collinear magnetic layers. Specifically, we demonstrate that the $30^{\circ}$-twisted honeycomb lattice bilayers form the dodecagonal quasicrystals supporting an $i$-wave altermagnetic phase. We further find that $45^{\circ}$-twisted bilayers preserving fourfold rotational symmetry support a mirror-driven $l$-wave phase with eight nodal lines. We characterize these altermagnetic patterns through plane-wave-projected spin-resolved spectral functions and spin-resolved conductances. Our work provides a systematic route to high-angular-order quasicrystalline altermagnetism through twist engineering.
\end{abstract}
	
\maketitle

\emph{Introduction.} 
Altermagnetism describes a class of collinear magnets in which a vanishing net magnetization coexists with momentum-dependent spin splitting without spin-orbit coupling~\cite{Smejkal2020sci,Ahn2019PRB,Naka2019NatCommun,Yuan2020PRB,Smejkal2022PRXissue3,Smejkal2022PRXissue4,Hayami2019JPSJ,Hirsch1990PRB,Mazin2021pnas,Ma2021NatCommun,McClarty2024PRL,Leeb2024PRL,Krempasky2024Nature,Li2024pnas,Jungwirth2026Nature}. In spin-degenerate antiferromagnets, symmetries such as combined spatial inversion and time reversal protect spin degeneracy. Altermagnets break these degeneracy-protecting symmetries but retain combined spin reversal and spatial rotation or reflection, which ensure zero net magnetization~\cite{Fender2025jacs,Liu2025NatPhys,Smejkal2022PRXissue3,Smejkal2022PRXissue4,Zeng2024PRB}. These retained symmetries constrain the characteristic momentum-dependent spin splitting~\cite{jungwirth2024arXiv,Bai2024AFM,Smejkal2022PRXissue3,Smejkal2022PRXissue4,Jungwirth2026Nature}. The resulting spin polarization can give rise to anisotropic spin transport~\cite{Naka2019NatCommun,GonzalezHernandez2021PRL,Bai2022PRL,Bai2023PRL,Lai2025PRL,Chen2025PRL,Wei2026PRB}. Altermagnetic signatures have been observed in both the momentum space and real space in various material systems~\cite{Krempasky2024Nature,Lee2024PRL,Osumi2024PRB,Reimers2024NatCommun,Ding2024PRL,Jiang2025NatPhys,Zhang2025NatPhys,Fu2025Altermagnet,Liu2024PRL,Yang2025NatCommun}. Recently, antiferromagnetism has been experimentally confirmed in the quasicrystal Au$_{56}$In$_{28.5}$Eu$_{15.5}$~\cite{Tamura2021jacs,Tamura2025NatPhys}. Subsequent theoretical studies have shown that quasicrystals can host altermagnetism~\cite{Li2026PRL,Chen2026PRL,Shao2026npjComputMater}, with spin-polarization patterns constrained by rotational symmetries forbidden in periodic crystals.

Twisted bilayers provide a platform for exploring the interplay between symmetry and magnetic order. Recent studies have shown that stacking and twisting magnetic layers can give rise to altermagnetism in periodic crystalline structures, with spin splitting exhibiting different angular patterns~\cite{He2023PRL,Pan2024PRL,Zeng2024PRBstackingaltermagnet,YCLiu2024PRL,Sheng2025SciChinaPhysMechAstron,Chen2026NatPhys}.
Meanwhile, twisting layers can also produce quasiperiodic structures known as moir\'e quasicrystals. Such structures have been explored in graphene and transition-metal dichalcogenide multilayers, with a prominent example being the experimentally realized dodecagonal quasicrystal in $30^{\circ}$-twisted bilayer graphene~\cite{Ahn2018sci,Yao2018pnas,Uri2023Nature,Hao2024NatCommun,Li2024Nature,Tsang2024Science,Liu2026NatPhys,Can2021NatPhys,Wang2020PRL,Xie2023NatPhys}. These advances motivate us to explore altermagnetism arising from the interplay between quasicrystalline symmetry and magnetic configurations in moir\'e quasicrystals.

In this work, we propose twisted magnetic bilayers as a platform for quasicrystalline altermagnetism. A $30^{\circ}$-twisted honeycomb bilayer with opposite ferromagnetic layer magnetizations supports rotation-driven $i$-wave altermagnetism. We further find that $45^{\circ}$-twisted bilayers formed from square lattices of spin clusters support mirror-driven $l$-wave altermagnetism. Reversing the spins in one layer of the twisted $l$-wave system could induce a rotation-driven $g$-wave phase. The emergent altermagnetism in these quasicrystals is characterized by alternating spin-resolved spectral functions. Moreover, we use spin-group symmetry to classify the altermagnetic phases in these bilayers. The classification shows how the relative twist, intralayer magnetic order, and interlayer magnetic alignment determine spin polarization. Furthermore, we adopt a two-tip transport measurement to probe the anisotropic spin transport of these altermagnetic phases. Our work establishes a connection between twisted quasicrystals and altermagnetism, offering a different route to designing quasicrystalline altermagnets.

\begin{figure}[t]
    \centering
    \includegraphics[width=1\linewidth]{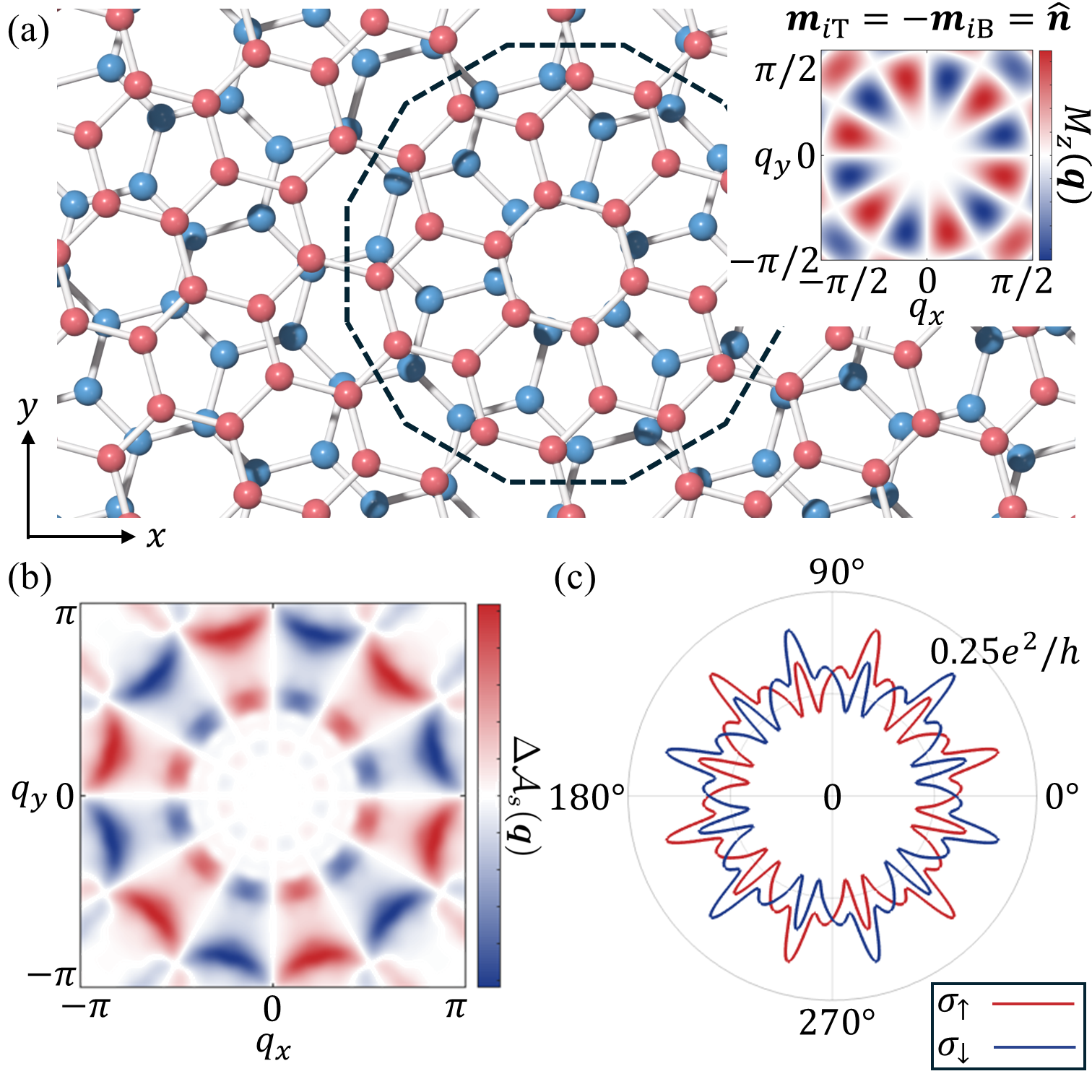}
    \caption{(a) Schematic of a $30^{\circ}$-twisted honeycomb bilayer with dodecagonal symmetry. Red and blue spheres denote opposite local spin magnetic moments. The inset shows the Fourier-transformed magnetic-moment distribution $M_z(\bm{q})$. (b) Spin-resolved spectral function difference $\Delta{\cal A}_s(\bm{q})={\cal A}_{\uparrow}(\bm{q})-{\cal A}_{\downarrow}(\bm{q})$ exhibiting the $i$-wave pattern. (c) Spin-resolved conductances $\sigma_{\uparrow}(\phi)$ and $\sigma_{\downarrow}(\phi)$. The parameters are $t=-1$, $t_{\perp}^{0}=3t$, $J_{\rm M}=-0.5t$, $a=1$, $d=1.2a$, $\xi=0.2a$, and $\eta=0.02|t|$. We use $n_F=0.1$ in (b) and $n_F=0.3$ with $R=2.5a$ in (c).}
    \label{fig1}
\end{figure}

\emph{Magnetic structure and electronic model.} 
We construct moir\'e quasicrystalline bilayers by twisting two magnetic layers. We first focus on a $30^{\circ}$-twisted honeycomb bilayer with opposite ferromagnetic layer magnetizations, as shown in Fig.~\ref{fig1}(a). This quasicrystal has dodecagonal symmetry. We describe the system using the spin-conserving Hamiltonian
\begin{align}
    H =  & - t \sum_{\langle ij \rangle,\alpha,s} c_{i\alpha s}^{\dagger} c_{j\alpha s} + \sum_{i,j,\alpha,\overline{\alpha},s} t_{\perp}(\bm{r}_{ij}) c_{i\alpha s}^{\dagger} c_{j\overline{\alpha} s} \\ \nonumber
    & - \mu \sum_{i,\alpha,s} c_{i\alpha s}^{\dagger} c_{i\alpha s} + J_{\text{M}} \sum_{i,\alpha,s,\overline{s}} c_{i\alpha s}^{\dagger} ( \bm{m}_{i\alpha} \cdot \bm{\sigma} )_{s\overline{s}} c_{ i\alpha \overline{s}} ,
    \label{hamFM}
\end{align}
where $i$ labels the lattice site and $s$ labels the spin. The layer index $\alpha=\text{T},\text{B}$ denotes the top and bottom layers, respectively. Here, $t$ is the intralayer nearest-neighbor hopping, $\mu$ is the chemical potential, $J_{\text{M}}$ parameterizes the local magnetization, and $\bm{\sigma}=(\sigma_x,\sigma_y,\sigma_z)$ is the vector of Pauli matrices acting in spin space. The interlayer hopping is given by $t_{\perp}(\bm{r}_{ij})=t_{\perp}^{0} e^{-(|\bm{r}_{ij}|-d)/\xi}$, where $\bm{r}_{ij}=\bm{r}_{i\alpha}-\bm{r}_{j \overline{\alpha}}$ is the relative displacement between sites $i,j$ in different layers, $d=|\bm{r}_{ij,\perp}|$ is the interlayer separation, and $\xi$ is the decay length. $\bm{m}_{i\alpha}$ is the local spin magnetic moment at site $\bm{r}_{i\alpha}$, and the magnetization directions are set to $\bm{m}_{i\text{T}}=\hat{\bm{n}}$ and $\bm{m}_{i\text{B}}=-\hat{\bm{n}}$. For the magnetic configuration in Fig.~\ref{fig1}(a), the N\'eel vector is set to $\hat{\bm{n}}=\hat{\bm{z}}$. We set the nearest-neighbor distance to $a=1$ and the chemical potential to $\mu=0$.

For the dodecagonal geometry, the nonmagnetic point group contains the twelvefold rotoinversion ${\cal S}_{12z}={\cal C}_{12z} {\cal P}$ and the twofold rotation ${\cal C}_{2v}$ about an in-plane axis $v$. Here, ${\cal C}_{12z}$ denotes a $30^{\circ}$ rotation about the $z$ axis and ${\cal P}$ denotes spatial inversion. With opposite layer magnetizations, the magnetic configuration is invariant under $[ {\cal C}_{2} || {\cal S}_{12z} ]$, where ${\cal C}_{2}$ is a twofold spin rotation about an axis perpendicular to the collinear magnetic moments. 

To visualize how the magnetic symmetry produces an alternating-sign pattern in momentum space, we examine the Fourier transform of the magnetic exchange field $J_{\mathrm M}\bm{m}_{i\alpha}$, which is defined as
\begin{equation}
    \bm{M}(\bm{q}) = \frac{1}{\sqrt{N}} \sum_{i,\alpha}
    \bm{m}_{i\alpha} e^{i\bm{q}\cdot\bm{r}_{i\alpha,\parallel}},
\end{equation}
where $\bm{q}$ is an in-plane wave vector, $\bm{r}_{i\alpha,\parallel}$ is the in-plane site position, and $N$ is the total number of sites in both layers. The inset of Fig.~\ref{fig1}(a) shows $M_z(\bm{q})$, which changes sign under a $30^{\circ}$ rotation, as required by the spin-group symmetry $[{\cal C}_{2}||{\cal S}_{12z}]$. The Fourier-transformed magnetic-moment distribution displays an $i$-wave pattern in momentum space, with positive and negative regions alternating every $30^{\circ}$.

The spin-polarized angle-resolved photoemission spectroscopy provides an experimental approach for detecting the proposed altermagnetic spin polarization~\cite{Ahn2018sci,Yao2018pnas,Rogalev2015NatCommun,Liu2024RuO2ARPES}. In the quasiperiodic system, the momentum-space texture of the spectral function can be characterized through a plane-wave projection of the real-space eigenstates. The spin-resolved spectral function is defined as~\cite{dOrnellas2026PRB}
\begin{align}
    {\cal A}_{s}(E_F,\bm{q}) = \sum_{n} { | \langle n_{\bm{r}} | \bm{q}, s \rangle | }^{2} \delta(E_F-E_n) ,
    \label{spectralfunction}
\end{align}
where $\delta(E_F-E_n)\simeq\tfrac{1}{\pi}\tfrac{\eta}{(E_F-E_n)^2+\eta^2}$ is the Lorentzian approximation to the Dirac delta function with broadening $\eta=0.02|t|$. The Fermi energy $E_F$ is determined by the filling number $n_{F}$. Here, $|n_{\bm{r}}\rangle$ denotes the real-space eigenstate with eigenvalue $E_n$. The $|\bm{q},s\rangle = \tfrac{1}{\sqrt{N}} \sum_{\bm{r}} e^{i \bm{q} \cdot \bm{r}} |\bm{r},s\rangle $ is the normalized plane-wave state in the spin-$s$ subspace. 

For the $30^{\circ}$-twisted honeycomb bilayer, the spin-resolved spectral function difference $\Delta {\cal A}_{s}(\bm{q}) = {\cal A}_{\uparrow}(\bm{q})-{\cal A}_{\downarrow}(\bm{q})$ changes sign under a $30^{\circ}$ rotation, as shown in Fig.~\ref{fig1}(b). This sign reversal identifies the $i$-wave altermagnetic phase with six nodal lines, which is protected by $[{\cal C}_{2}||{\cal S}_{12z}]$ symmetry. $M_z(\bm{q})$ exhibits an angle-dependent sign reversal pattern similar to that of $\Delta {\cal A}_{s}(\bm{q})$, as shown in the inset of Fig.~\ref{fig1}(a).

\emph{Spin-group classification.} 
The $30^{\circ}$-twisted honeycomb bilayer provides one example of quasicrystalline altermagnetism. We now develop a broader classification to determine which altermagnetic patterns can arise from different lattice symmetries and magnetic configurations. We describe the magnetic symmetries using spin-group operations~\cite{Lifshitz1998PRL,PFLiu2022PRX,Xiao2024PRX,Chen2024PRX,XBChen2025Nature,YTLiu2026Nature}. For two layers with ${\cal C}_{mz}$ symmetry ($m=2,3,4,6$), a relative twist of $\pi/m$ produces a bilayer with rotoinversion symmetry ${\cal S}_{2mz}={\cal C}_{2mz}{\cal P}$.
The cases $m=4$ and $m=6$ give octagonal and dodecagonal structures forbidden in translationally symmetric systems. 
The collinear magnetic configurations preserve the effective antiunitary symmetry $[{\cal C}_{2}||{\cal E}]{\cal T}$, which constrains the Fourier-transformed magnetic-moment distribution to satisfy the even-parity relation $M_z(\bm{q})=M_z(-\bm{q})$. Here, ${\cal E}$ and ${\cal T}$ denote the identity and time-reversal operations, respectively. With this even-parity constraint, $[{\cal C}_{2}||{\cal S}_{2mz}]$ requires a sign reversal under a $\pi/m$ rotation as $M_z({\cal C}_{2mz}\bm{q})=-M_z(\bm{q})$. When the magnetic configuration preserves $[{\cal E}||{\cal S}_{2mz}]$ and $[{\cal C}_{2}||{\cal C}_{2v}]$, they together impose $M_z(-{\cal C}_{2mz}{\cal C}_{2v}\bm{q})=-M_z(\bm{q})$. Here, the combined spatial operation ${\cal C}_{2mz} {\cal C}_{2v}$ acts as a reflection on the in-plane wave vector.

Table~\ref{table1} summarizes the spin-group symmetries and magnetic phases associated with different intralayer and interlayer magnetic orders. Except for configurations preserving $[{\cal C}_{2}||{\cal C}_{2z}]$, which remain spin-degenerate, the listed phases exhibit either rotation-driven or mirror-driven quasicrystalline altermagnetism. In the spin-degenerate cases, $[{\cal C}_{2}||{\cal C}_{2z}]$ reverses both the in-plane momentum and the spin, whereas $[{\cal C}_{2}||{\cal E}]{\cal T}$ reverses the momentum without changing the spin. Their combined action connects opposite-spin states at the same momentum and energy, enforcing ${\cal A}_{\uparrow}(E,\bm{q})={\cal A}_{\downarrow}(E,\bm{q})$.


\begin{table*}[!hbpt]
    \caption{Spin-group classification of collinear magnetic configurations in twisted moir\'e quasicrystals. The intralayer FM and AFM labels denote ferromagnetic and antiferromagnetic order within each layer. The interlayer FM and AFM labels indicate parallel and antiparallel alignment of the layer magnetizations or N\'eel orders, respectively. The sign reversals of spin polarization are counted over a $2\pi$ rotation in momentum-space.}
    \label{table1}
    \begin{ruledtabular}
    \tabcolsep=0.1cm
    \renewcommand{\arraystretch}{1.2}
    \begin{tabular}{ccccccc}
    \multirow{2}{*}{Twisted quasicrystal} & \multicolumn{2}{c}{Magnetic order} & \multirow{2}{*}{Decisive spin-group operations} & \multirow{2}{*}{Altermagnetism} & \multirow{2}{*}{Sign reversals} & \multirow{2}{*}{Mechanism} \\
    \cline{2-3}
    & Intralayer & Interlayer & & \\
    \hline
    \multirow{5}{*}{$12$-fold} & FM & AFM & $[{\cal C}_{2}||{\cal S}_{12z}]$ & $i$-wave & 12 & rotation-driven \\
    \cline{2-7}
    & \multirow{2}{*}{AFM} & \multirow{2}{*}{AFM} & $[{\cal C}_{2}||{\cal S}_{12z}]$ & $i$-wave & 12 & rotation-driven \\
    \cline{4-7}
    & & & $[{\cal C}_{2}||{\cal C}_{2z}]$ & \multicolumn{3}{c}{spin-degenerate antiferromagnetism} \\
    \cline{2-7}
    & \multirow{2}{*}{AFM} & \multirow{2}{*}{FM} & $[{\cal C}_{2}||{\cal C}_{2z}]$ & \multicolumn{3}{c}{spin-degenerate antiferromagnetism} \\
    \cline{4-7}
    & & & $[{\cal E}||{\cal S}_{12z}]$ and $[{\cal C}_{2}||{\cal C}_{2v}]$ & $q$-wave & 24 & mirror-driven \\
    \hline
    \multirow{5}{*}{$8$-fold} & FM & AFM & $[{\cal C}_{2}||{\cal S}_{8z}]$ & $g$-wave & 8 & rotation-driven \\
    \cline{2-7}
    & \multirow{2}{*}{AFM} & \multirow{2}{*}{AFM} & $[{\cal C}_2||{\cal S}_{8z}]$ & $g$-wave & 8 & rotation-driven \\
    \cline{4-7}
    & & & $[{\cal C}_2||{\cal C}_{4z}]$ & $d$-wave & 4 & rotation-driven \\
    \cline{2-7}
    & \multirow{2}{*}{AFM} & \multirow{2}{*}{FM} & $[{\cal C}_2||{\cal C}_{4z}]$ & $d$-wave & 4 & rotation-driven \\
    \cline{4-7}
    & & & $[{\cal E}||{\cal S}_{8z}]$ and $[{\cal C}_2||{\cal C}_{2v}]$ & $l$-wave & 16 & mirror-driven
    \end{tabular}
    \end{ruledtabular}
\end{table*}

\begin{figure*}[!hbtp]
    \centering
    \includegraphics[width=0.8\linewidth]{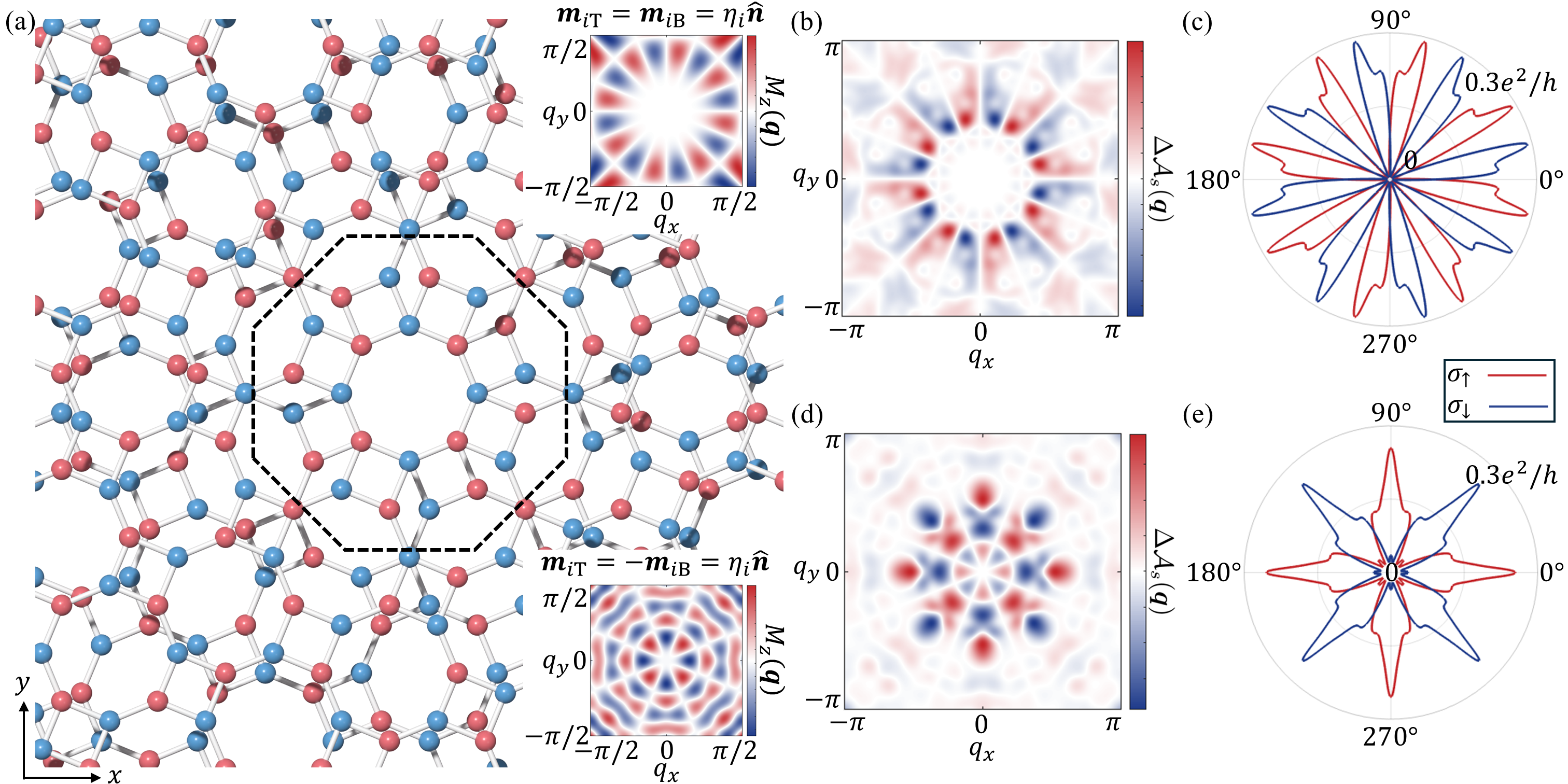}
    \caption{(a) Schematic illustration of the $45^{\circ}$-twisted lattice structure. Red and blue spheres denote opposite local spin magnetic moments. The upper-right and lower-right insets show the Fourier-transformed magnetic-moment distributions $M_z(\bm{q})$ for the $l$-wave and $g$-wave phases, respectively. (b) Spin-resolved spectral function difference $\Delta{\cal A}_{s}(\bm{q})$ for the mirror-driven $l$-wave phase. (c) Spin-resolved conductances $\sigma_{\uparrow}(\phi)$ and $\sigma_{\downarrow}(\phi)$ for the $l$-wave phase. (d) Spin-resolved spectral function difference $\Delta{\cal A}_{s}(\bm{q})$ for the rotation-driven $g$-wave phase. (e) Spin-resolved conductances $\sigma_{\uparrow}(\phi)$ and $\sigma_{\downarrow}(\phi)$ for the $g$-wave phase. Parameters in (b) and (d) are the same as those in Fig.~\ref{fig1}(b). Parameters in (c) and (e) are the same as those in Fig.~\ref{fig1}(c), except that $n_F=0.465$ and $0.447$, respectively.}
    \label{fig2}
\end{figure*}

This classification in Table~\ref{table1} identifies the $30^{\circ}$-twisted honeycomb bilayer as a rotation-driven $i$-wave altermagnet and predicts a mirror-driven $l$-wave phase in a $45^{\circ}$-twisted bilayer preserving $[{\cal E}||{\cal C}_{4z}]$ symmetry. Reversing all spins in one layer of the $45^{\circ}$-twisted square-octagon bilayer switches the mirror-driven $l$-wave phase to the rotation-driven $g$-wave phase. This reversal replaces the spin-preserving operation $[{\cal E}||{\cal S}_{8z}]$ with the spin-flipping operation $[{\cal C}_{2}||{\cal S}_{8z}]$. The classification further predicts mirror-driven $q$-wave altermagnetism in dodecagonal quasicrystals, as demonstrated in the Supplemental Material~\cite{ourSM}.

\emph{Magnetic order controlled quasicrystalline altermagnetism.} 
The symmetry classification in Table~\ref{table1} allows mirror-driven $l$-wave altermagnetism in antiferromagnetic layers. We consider a square-octagon lattice of four-site square clusters with checkerboard antiferromagnetic order~\cite{Ye2011PRL,Bao2011CPL,Zhu2026PRL}. The local spin magnetic moments are set to $\bm{m}_{i\text{T}}=\bm{m}_{i\text{B}}=\eta_i \hat{\bm{n}}$, where $\eta_i=\pm 1$ labels the antiferromagnetic order, as illustrated in Fig.~\ref{fig2}(a). The $45^{\circ}$-twisted bilayer with identical antiferromagnetic order in both layers preserves $[{\cal E}||{\cal S}_{8z}]$ and $[{\cal C}_{2}||{\cal C}_{2x}]$ symmetries. These symmetries protect an $l$-wave pattern with eight symmetry-enforced nodal lines. The reflection-induced sign reversal is manifested in the Fourier-transformed magnetic-moment distribution $M_z(\bm{q})$ in the upper-right inset of Fig.~\ref{fig2}(a) and the spin-resolved spectral function difference $\Delta{\cal A}_s(\bm{q})$ in Fig.~\ref{fig2}(b). The conductance difference $\Delta \sigma_s(\phi)$ exhibits the same reflection-induced sign reversal, as shown in Fig.~\ref{fig2}(c).

By reversing the N\'eel order in the bottom layer, we switch from mirror-driven $l$-wave to rotation-driven $g$-wave altermagnetism within the same lattice structure. The local spin magnetic moments in the $g$-wave phase now satisfy $\bm{m}_{i\text{T}}=-\bm{m}_{i\text{B}}=\eta_i \hat{\bm{n}}$. In this case, the spin-preserving operation $[{\cal E}||{\cal S}_{8z}]$ is replaced by the spin-flipping operation $[{\cal C}_{2}||{\cal S}_{8z}]$. The lower-right inset of Fig.~\ref{fig2}(a) shows the corresponding $M_z(\bm{q})$. The spin-resolved spectral function difference exhibits a $g$-wave pattern, as shown in Fig.~\ref{fig2}(d). Figure~\ref{fig2}(e) shows that the spin-resolved conductances are interchanged when the detection angle $\phi$ is increased by $\pi/4$. Other intralayer magnetic configurations and the corresponding altermagnetic symmetry classes are presented in the Supplemental Material~\cite{ourSM}.

\emph{Spin-resolved transport.} 
The spin-resolved spectral function difference reveals the angular dependence of the spin polarization. To probe this angular dependence in transport, we consider two contacts at opposite positions $\pm R(\cos\phi,\sin\phi)$ on the sample. Here, $\phi$ is the polar angle and $R$ is the radial distance from the sample center. We evaluate the spin-resolved conductances $\sigma_{\uparrow}(\phi)$ and $\sigma_{\downarrow}(\phi)$ as functions of the contact angle $\phi$. Their difference $\Delta\sigma_{s}(\phi)=\sigma_{\uparrow}(\phi)-\sigma_{\downarrow}(\phi)$ can exhibit angle-dependent sign changes constrained by the spin-group symmetry, and provide a transport signature of altermagnetism.

We calculate the spin-resolved conductances $\sigma_{s}(\phi)$ using the Landauer-B\"uttiker formula~\cite{RolfLandauer1970,Buettiker1988PRB,Fisher1981PRB,MacKinnon1985,Metalidis2005PRB}. At zero temperature, the linear conductance in spin channel $s$ is $\sigma_{s}=\tfrac{e^2}{h} \operatorname{Tr}[ \Gamma_{\text{in},s} G^{R}_{s} \Gamma_{\text{out},s} G^{A}_{s} ]$. Here, $G^{R}_{s}$ and $G^{A}_{s}$ are the retarded and advanced Green functions of the quasicrystal coupled to the contacts. The contact broadening matrices are defined by $
\Gamma_{\ell,s}=i(\Sigma_{\ell,s}^{R}
-\Sigma_{\ell,s}^{A})$ for $\ell=\mathrm{in},\mathrm{out}$, where $\Sigma_{\ell,s}^{{R},{A}}$ are the corresponding contact self-energies. For the $30^{\circ}$-twisted honeycomb bilayer, the $[ {\cal C}_{2} || {\cal S}_{12z} ]$ symmetry exchanges the two spin-resolved conductances $\sigma_{s}(\phi)$ when $\phi$ increases by $\pi/6$, as shown in Fig.~\ref{fig1}(c). Thus, the conductance difference $\Delta\sigma_s(\phi)$ changes sign upon a $30^{\circ}$ rotation of the contact pair, providing a transport signature of the $i$-wave altermagnetic phase. The spin-resolved conductances for the $l$- and $g$-wave phases are shown in Figs.~\ref{fig2}(c) and \ref{fig2}(e), respectively.

\emph{Conclusion and discussion.} 
We have demonstrated quasicrystalline altermagnetism in twisted magnetic bilayers governed by the interplay between lattice symmetry and magnetic order. Based on spin-group symmetry, we classify the possible magnetic configurations into spin-degenerate and altermagnetic classes, with the latter including rotation-driven and mirror-driven phases. The $30^{\circ}$-twisted honeycomb bilayer exhibits $i$-wave altermagnetism protected by the spin-flipping symmetry $[ {\cal C}_{2} || {\cal S}_{12z} ]$. For the $45^{\circ}$-twisted square–octagon bilayer, reversing the N\'eel order in one layer switches mirror-driven $l$-wave to rotation-driven $g$-wave altermagnetism. The angle-dependent spin-resolved conductances further reveal the altermagnetic transport signatures. These results establish twist engineering as a promising route to realizing and controlling quasicrystalline altermagnetism.

Layered magnetic materials provide potential platforms for realizing quasicrystalline altermagnetism. For rotation-driven altermagnetism in bilayers with opposite ferromagnetic layer magnetizations, CrI$_3$ provides ferromagnetic layers with stacking-dependent magnetism~\cite{Song2021sci,Xu2022NatNanotechnol,Xie2022NatPhys,Li2024CrI3,Wong2026NatNano}, while FeCl$_2$ exhibits intralayer ferromagnetism and interlayer antiferromagnetic coupling~\cite{Wilkinson1959PR,Zhou2024ACSNano}. For mirror-driven altermagnetism in bilayers with intralayer antiferromagnetic order, vacancy-ordered iron selenides exhibit block-antiferromagnetic order and may be suitable for realizing the square-octagon model~\cite{Ye2011PRL,Bao2011CPL,XMa2026PRB}. Recent experimental evidence of altermagnetism in orthogonally twisted CrPS$_4$ homostructures further supports the possibility of realizing quasicrystalline altermagnetism through twisting~\cite{Chen2026NatPhys}. 

\emph{Acknowledgments.} 
This work was supported by the National Natural Science Foundation of China (NSFC, Grants No. 92565103, No. 12474151, No. 12547101, No. 12304195, No.12674198, and No.125B2066), the Natural Science Foundation of Chongqing (Grant No. CSTB2025NSCQLZX0010), Beijing National Laboratory for Condensed Matter Physics (No. 2024BNLCMPKF025), and the Fundamental Research Funds for the Central Universities (Grant No. 2025CDJIAISYB-032). R.C. acknowledges the Chutian Scholars Program in Hubei Province, the Hubei Provincial Natural Science Foundation (Grant No. 2025AFA081), the key project of Hubei provincial department of education (under Grant No. D20241004), and the original seed program of Hubei University.

\bibliography{references}

\end{document}